\documentclass[epj,twocolumn,english]{webofc}
\usepackage[varg]{txfonts}   
\woctitle{LHCP 2013}

\newcommand{\ttbar}{\ensuremath{\rm{} t \bar{t}}}
\newcommand{\Alep}{\ensuremath{\rm{} A_{FB}^{lep}}}
\newcommand{\AFB}{\ensuremath{\rm{} A_{FB}}}

\begin{document}
\title{Leptonic asymmetry in \ttbar{} production at CDF}
%
%

\author{Stefano Camarda\inst{1}\fnsep\thanks{\email{stefano.camarda@desy.de}}
}

\institute{DESY - Hamburg
          }

\abstract{
The leptonic asymmetry in semileptonic \ttbar{} decays is measured with the CDF detector using the full Tevatron Run II dataset,
which corresponds to $9.4~\rm fb^{-1}$ of integrated luminosity. The measured asymmetry is extrapolated to the full
kinematic range and the measured value of $\Alep = 0.094^{+0.032}_{-0.029}$ is compared to the NLO prediction $\Alep = 0.038 \pm 0.003$.
}
\maketitle
\section{Introduction}
\label{intro}
The CDF and D0 experiments have measured the forward-backward asymmetry \AFB{} for \ttbar{} production
in $\rm{} p \bar{p}$ collisions \cite{Aaltonen:2011kc,Abazov:2011rq}, the CDF measurement reports
$\AFB = 0.164 \pm 0.045$, the D0 measurement $\AFB = 0.196 \pm 0.065$.
Both results are higher than the prediction $\AFB = 0.088 \pm 0.006$ \cite{Bernreuther:2012sx},
which includes both electroweak and QCD next-to-leading order (NLO) corrections.
Much effort has been invested to improve the theoretical calculation of the asymmetry in the Standard Model,
through the estimation of beyond NLO corrections and related uncertainties. Soft gluon resummation was found
to give a negligible contribution \cite{Ahrens:2011uf}, electroweak corrections are of the order of $25\%$ and are
included in the current predictions \cite{Hollik:2011ps,Bernreuther:2012sx},
a calculation based on the 
Principle of Maximum Conformality (PMC) for the scale setting reports a $40\%$ enhancement, and finally
a more realistic estimate of scale uncertainties at NLO of the order $30\%$ should be considered \cite{Bernreuther:2012sx,Campbell:2012uf,Brodsky:2012ik}.

The leptonic asymmetry, defined as
\begin{equation}
\label{eq:alep}
\Alep = \frac{N(qy_l > 0) - N(qy_l < 0)}{N(qy_l > 0) + N(qy_l < 0)}
\end{equation}
where $\rm{} q$ is the lepton charge and $\rm{} y_l$ the lepton rapidity in the
laboratory frame, is an observable related to \AFB{} which can provide complementary information.
The measurement of \Alep{} depends only on lepton charge and direction,
and therefore can be measured very precisely.
The D0 experiment reported measurements of the leptonic asymmetry in \ttbar{} production both in the semileptonic and 
in the di-lepton channels with about half of the full Tevatron Run II dataset, 
the combined result is $\Alep = 0.118 \pm 0.032$ \cite{Abazov:2012oxa}.

There are two physical origins of leptonic asymmetry \Alep{}, the \AFB{} asymmetry,
and the polarisation of the \ttbar{} system.
Leptons partially inherit the asymmetry of the parent tops, and in addition 
the V-A coupling of the weak interaction connects the direction of the top decay products to the polarisation of the top quarks.
Top pairs are produced unpolarised in the Standard Model, an excess of right handed top pairs would enhance \Alep{}, 
while left-handed pairs would induce a negative contribution.

The relationship between the top asymmetry, the \ttbar{} polarisation and the leptonic asymmetry has been the subject of
recent theoretical work in the context of possible explanations of the top asymmetry \AFB{} \cite{Falkowski:2012cu,Berger:2012tj}.

\section{Physics models and expected asymmetry}
\label{sec:physics-models}

In the measurement of \Alep{} several reference models and corresponding Monte Carlo samples are used, they are listed in table
\ref{tab:samples}.

\begin{table*}
\centering
\caption{Reference models used for the \Alep{} measurement.}
\label{tab:samples}       
\begin{tabular}{lllll}
\hline
Model & \AFB{} & \Alep{} & Polarisation & \\
\hline
\hline
\textsc{alpgen} & 0.000(1)  & +0.003(1) & +0.009(2) & LO Standard Model \\
\textsc{powheg} & 0.052(0)  & +0.024(1) & +0.001(1) & NLO Standard Model \\
OCTET A         & +0.156(1) & +0.070(2) & -0.005(3) & unpolarised axigluon \\
OCTET L         & +0.121(1) & -0.062(1) & -0.290(3) & left-handed axigluon\\
OCTET R         & +0.114(2) & +0.149(2) & +0.280(3) & right-handed axigluon \\
\hline
\end{tabular}
\end{table*}

All the Monte Carlo samples are showered with \textsc{pythia} \cite{Sjostrand:2006za} and processed with the full CDF detector simulation.
\textsc{alpgen} \cite{Mangano:2002ea} is a LO matrix-element matched to PS generator which predicts no asymmetry, \textsc{powheg} \cite{Frixione:2007nw}
is a NLO generator which predicts a small asymmetry, OCTET A, L and R are axigluon models simulated with
\textsc{madgraph} \cite{Alwall:2007st} which predict \AFB{} comparable to the measured values, but different \ttbar{} polarisation 
and different values of \Alep{}.
The two polarised models, Octet L and Octet R, are light ($200~\rm{} GeV/c^2$) and wide ($50~\rm{} GeV/c^2$) axigluons.
Octet L has a left-handed coupling and negative polarisation, while Octet R has a right-handed
coupling and positive polarisation. Octet A is a massive ($2.0~\rm{} TeV/c^2$) and narrow axigluon with
unpolarised couplings.

A Standard Model NLO QCD fixed order calculation of \Alep{} including electroweak corrections reports $0.038 \pm 0.003$ \cite{Bernreuther:2012sx}.
When comparing the NLO fixed order result to the prediction from a NLO generator interfaced to parton shower,
an important difference has to be considered.
The first non trivial orders for the numerator and denominator of equation (\ref{eq:alep}) are respectively $\rm{} O(\alpha_s^3)$ and $\rm{} O(\alpha_s^2)$,
for this reason in the fixed order calculation the $\rm{} O(\alpha_s^2)$ result of the inclusive \ttbar{} cross section
is used in the denominator of the \Alep{} asymmetry.
In a NLO Monte Carlo generator like \textsc{powheg}, both numerator and denominator of equation (\ref{eq:alep}) are computed at the same $\rm{} O(\alpha_s^3)$ order, which leads
to a sizeable difference with respect to the fixed order result.
The \textsc{powheg} prediction also does not include the electroweak corrections, which enhance \Alep{} of $26\%$.

A data-driven estimation of \Alep{} within the Standard Model can be done dividing the \AFB{} measured at CDF by the ratio $\frac{\AFB{}}{\Alep{}} = 2.17$
as predicted by \textsc{powheg}. With this assumption the expected \Alep{} is $0.076$.

\section{Event selection ad sample composition}
\label{sec:sample}
The full CDF Run II dataset, corresponding to an integrated luminosity of $9.4~\rm{} fb^{-1}$ is used to measure \Alep{}.
The \ttbar{} semileptonic events are selected with high-$\rm{} p_T$
electron or muon and large missing $\rm{} E_T$ triggers.
Jets are reconstructed with the jetclu cone algorithm in a radius $\rm{} R = 0.4$.
Events are selected with exactly one lepton with $\rm{} p_T > 20$ GeV/c and $\rm{} |y_l| < 1.25$, missing $\rm{} E_T > 20$ GeV, at least 4 jets with $|\eta| < 2.0$,
at least 3 jets with $\rm{} E_T > 20$ GeV, at least one jet with $\rm{} E_T > 12$ GeV, at least 1 b-tagged jet, and $\rm{} H_T > 220$ GeV.
After the event selection the sample is mainly compose by \ttbar{} events, with the main background coming from W+jets events.

\begin{figure}
\centering
\includegraphics[width=0.45\textwidth,clip]{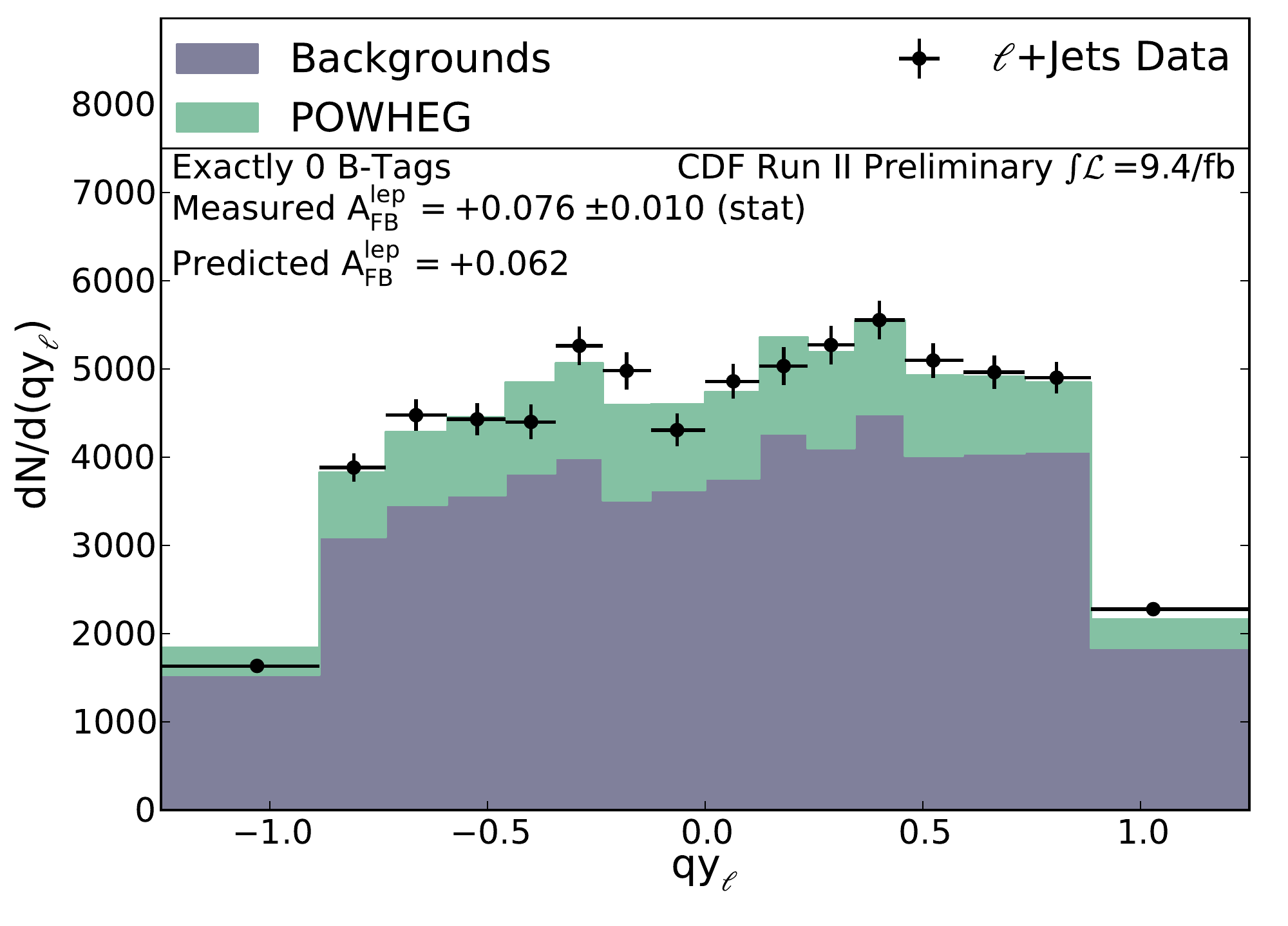}
\caption{$\rm{} qy_l$ distribution in the background-enhanced control region.}
\label{fig:background}       
\end{figure}

Background processes are expected to contribute a nonzero asymmetry, in particular the largest background,
namely W+jets, is asymmetric for a combination of electroweak and PDF effects. 
In order to validate the modelling of the leptonic asymmetry in the background simulation
a background-enhanced control region is defined requiring that none of the jets is identified as a b-tagged jet.
Figure \ref{fig:background} shows the signed rapidity distribution $\rm{} qy_l$ in the control region,
where the W+jets background is simulated with the \textsc{alpgen+pythia} Monte Carlo.
The observed leptonic asymmetry of $0.076$ in the background-enhanced region
is in good agreement with the expected value of $0.062$.

\section{Extrapolation to the full kinematic region}
\label{sec:extrapolation}
\begin{figure*}
\centering
\includegraphics[width=0.49\textwidth,clip]{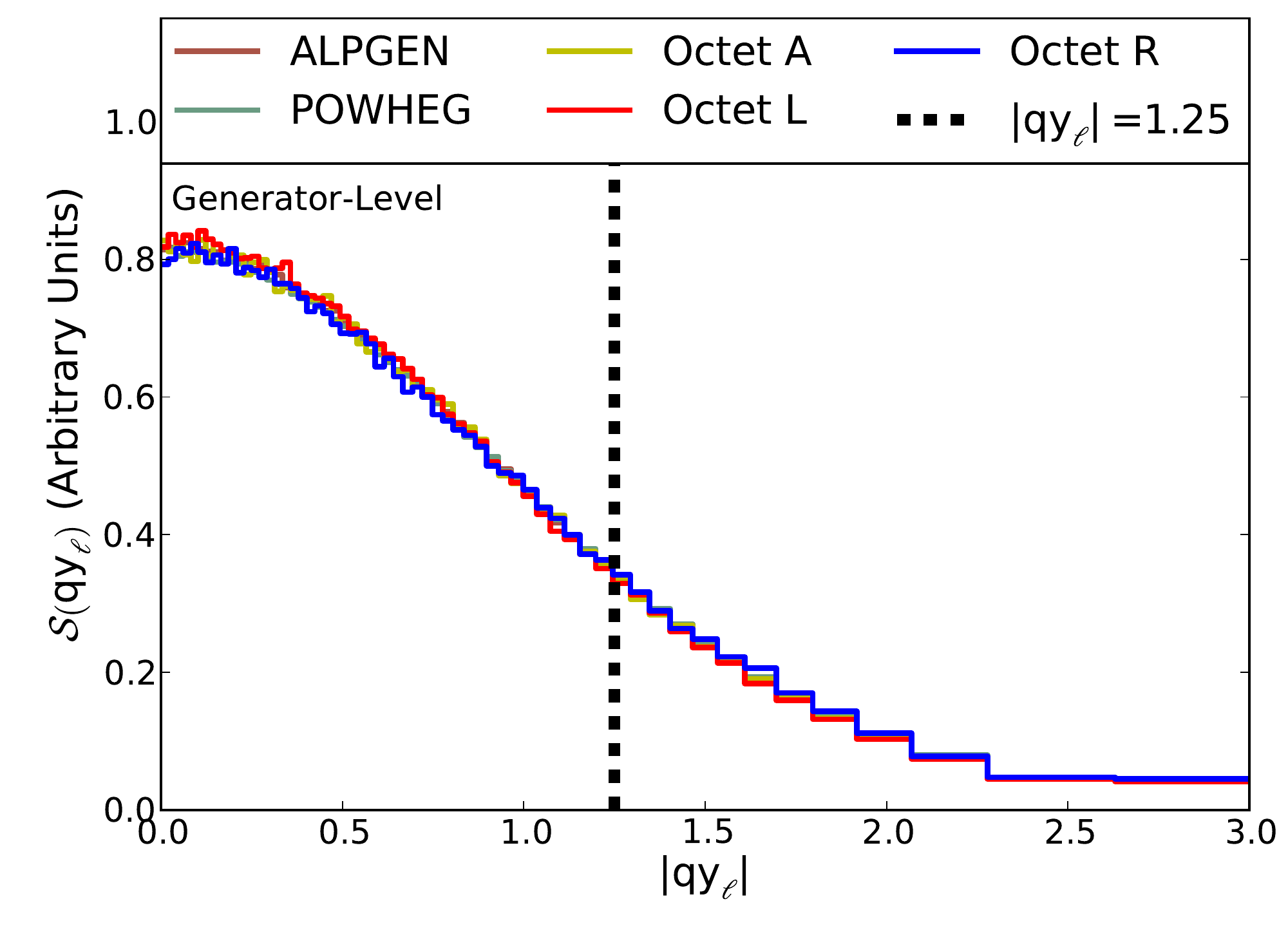}
\includegraphics[width=0.49\textwidth,clip]{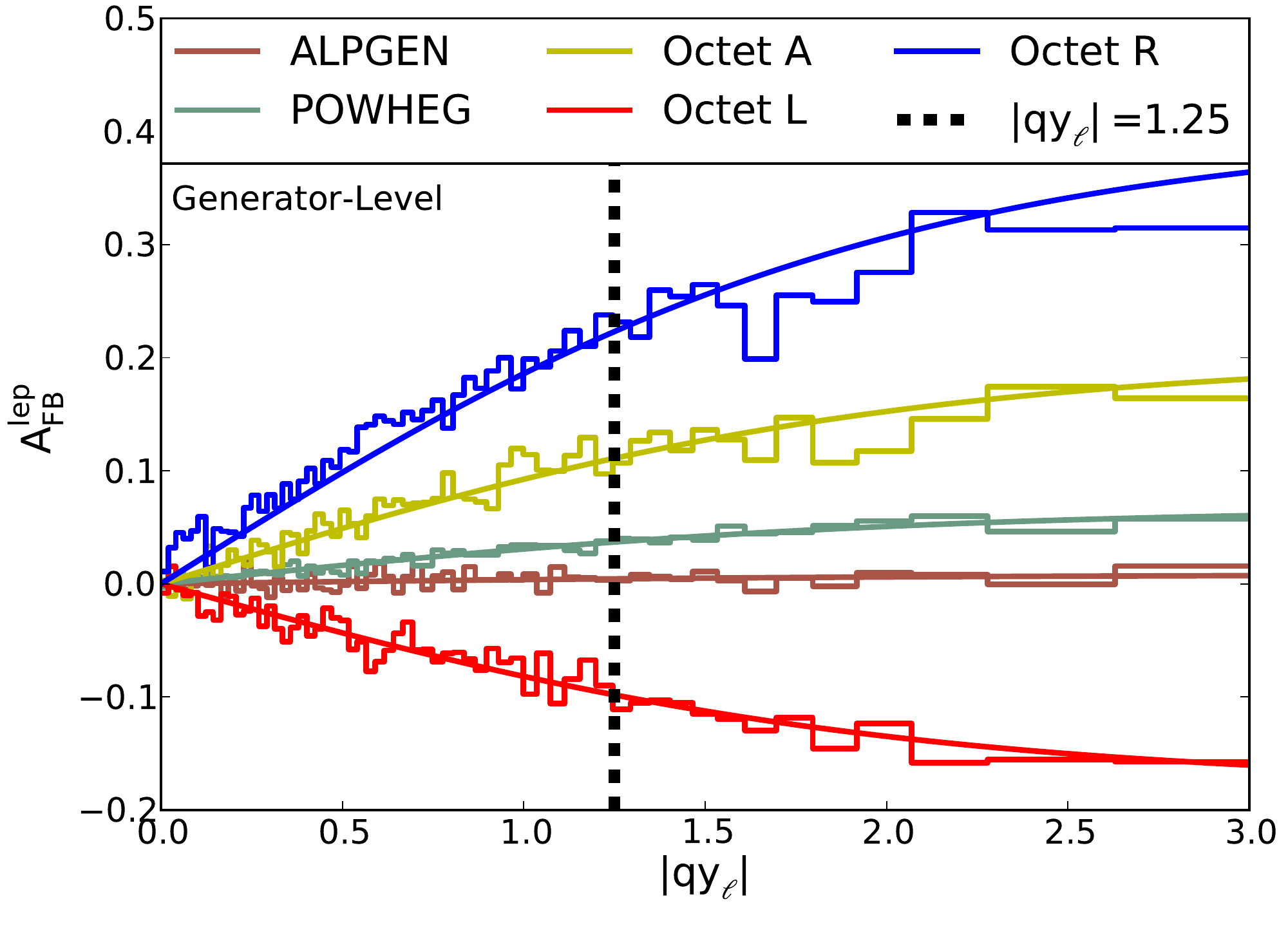}
\caption{Symmetric and asymmetric components $\rm{} S(qy_l)$ and $\rm{} A(qy_l)$ of the signed rapidity distribution $\rm{} qy_l$ in the reference models.}
\label{fig:asymdec}       
\end{figure*}

The signed rapidity distribution $\rm{} qy_l$ is measured in the limited range $\rm{} |y_l| < 1.25$, which corresponds to the
detector acceptance. In order to extrapolate the measurement to the full kinematic space,
$\rm{} N(qy_l)$ is decomposed into symmetric and asymmetric components:
\begin{eqnarray}
\rm{} S(qy_l) = \frac{N(qy_l) + N(-qy_l)}{2} \\
\rm{} A(qy_l) = \frac{N(qy_l) - N(-qy_l)}{N(qy_l) + N(-qy_l)}
\end{eqnarray}
The symmetric part is the same in all the considered models of table \ref{tab:samples}, while the asymmetric part 
capture the difference between the models, as shown in figure \ref{fig:asymdec}.

The measured $\rm{} A(qy_l)$ is unfolded to the parton level, fitted with a hyperbolic tangent function
\begin{equation}
\label{eq:fit}
\rm{} F(qy_l) = a \cdot tanh[\frac{1}{2}qy_l]
\end{equation}
and convoluted with the symmetric component $\rm{} S(qy_l)$ evaluated with the \textsc{powheg} Monte Carlo
 to extract the measured leptonic asymmetry:
\begin{equation}
\label{eq:conv}
\Alep{} = \frac{\int_{0}^{\infty}dqy_l[A(qy_l) \times  S(qy_l)]} {\int_{0}^{\infty}dqy_l S(qy_l)}
\end{equation}

\section{Results}
\label{sec:Results}
The largest systematic uncertainty on \Alep{} is associated to the background subtraction.
The background uncertainty is evaluated with a pseudo-experiment technique which accounts simultaneously for the uncertainty
on the normalisation of the backgrounds and for the uncertainty on the shape due to limited statistics
of the Monte Carlo samples.

Another important source of uncertainty comes from the modelling of the \ttbar{}
recoil due to QCD radiation.
The presence of radiated jets is strongly correlated with both \AFB{} and the $\rm{} p_T$ of
the \ttbar{} system. 
Colour predominantly flows from an initiating light quark to an outgoing
top-quark and from an anti-quark to an anti-top.
As a consequence, events with larger difference between initial state quark and top directions
are associated with harder QCD radiation.
Events with more radiation have a larger acceptance because can more easily pass the high $\rm{} p_T$ selection requirements.
The uncertainty on the recoil modelling is estimated
comparing the acceptance of the nominal \textsc{powheg} Monte Carlo
with two other models, namely \textsc{pythia}
and \textsc{alpgen+pythia}.
The recoil spectra of both \textsc{pythia} and \textsc{alpgen+pythia}
are harder than \textsc{powheg} and give larger results for \Alep{}, the 
uncertainty is therefore single-sided.
An additional uncertainty related to the recoil model may arise from the
initial-state radiation model in the \textsc{pythia} parton shower.
The uncertainty is estimated performing variations of the initial and final state radiation parameters (IFSR),
the effect is found to be small.

Other QCD and jets related sources of uncertainties like colour reconnection, parton shower model, and jet-energy-scale,
have been estimated.
They give a small contribution to the \Alep{} unertainty because hadronic jets are used in the measurement only to select the event sample.
PDF uncertainties largely cancel between the numerator and denominator in the definition of \Alep{}.
The uncertainties on \Alep{} are listed in table \ref{tab:uncertainties}.

\begin{table}
\centering
\caption{Uncertainties of the \Alep{} measurement.}
\label{tab:uncertainties}       
\begin{tabular}{ll}
\hline
Backgrounds & 0.015 \\
Recoil modelling & $^{+0.013}_{-0.000}$\\
Colour reconnection & 0.0067 \\
Parton shower & 0.0027 \\
PDF & 0.0025 \\
Jet-energy-scale & 0.0022 \\
IFSR & 0.0018 \\
\hline
Total systematic & $^{+0.022}_{-0.017}$\\
\hline
Statistical uncertainty & 0.024 \\
\hline
\hline
Total uncertainty & $^{+0.032}_{-0.029}$\\
\hline
\end{tabular}
\end{table}

\begin{figure}
\centering
\includegraphics[width=0.45\textwidth,clip]{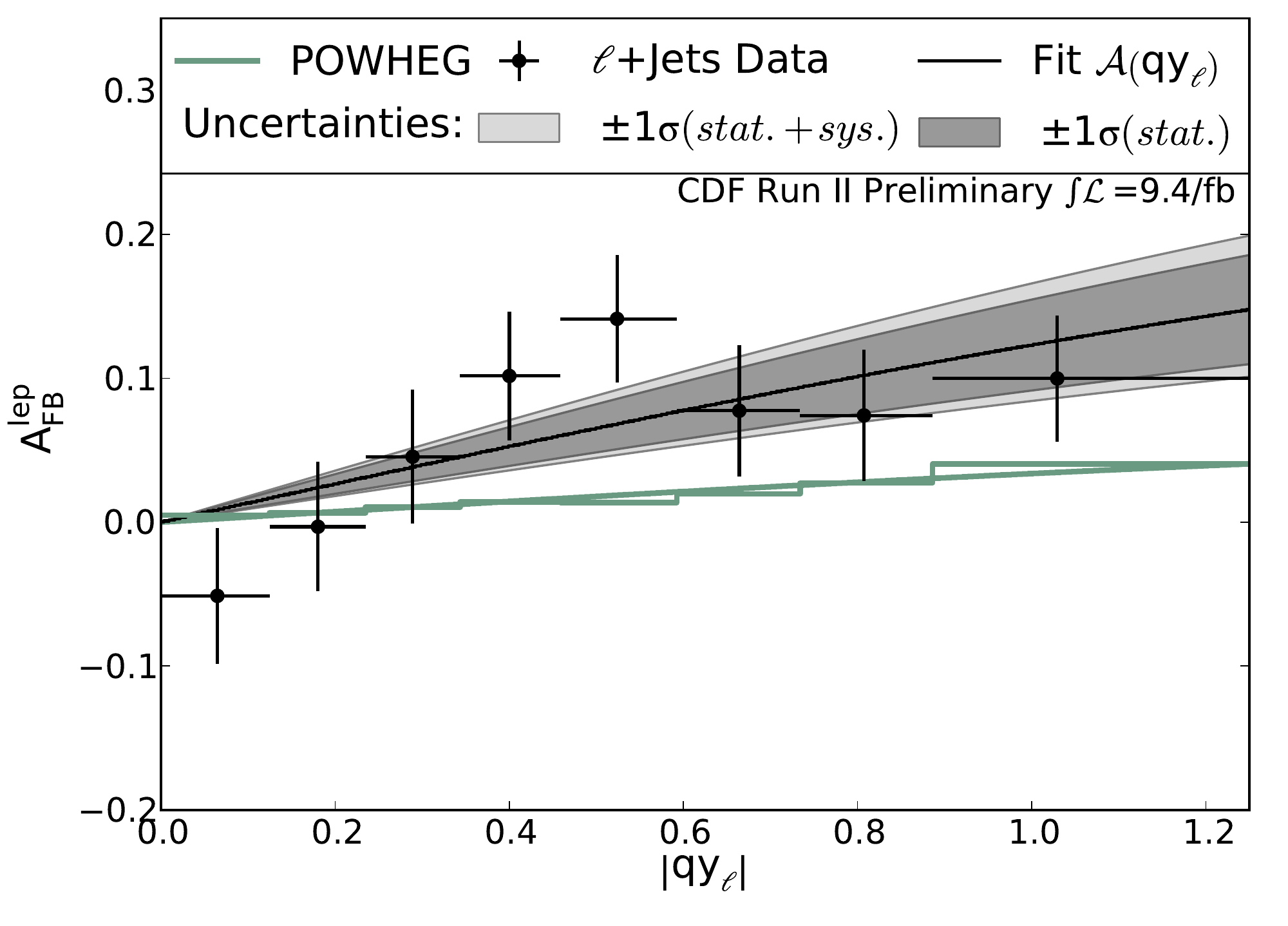}
\caption{Asymmetric component $\rm{} A(qy_l)$ of the signed rapidity distribution $\rm{} qy_l$ as measured in the data (black points) and compared to the \textsc{powheg} prediction (green).
A hyperbolic tangent fit to the data and to the prediction is shown as a smooth curve of same colours.
The dark (light) grey bands shows the statistical (total) uncertainty on the fit result.
}
\label{fig:asymdata}       
\end{figure}

Figure \ref{fig:asymdata} shows the parton level unfolded $\rm{} A(qy_l)$ measured in the data, compared
to the \textsc{powheg} prediction, and the result of fits to both data and Monte Carlo with 
equation (\ref{eq:fit}). After the convolution of equation (\ref{eq:conv}) with $\rm{} S(qy_l)$ as estimated from \textsc{powheg},
the measured parton level leptonic asymmetry in the full kinematic region is $\Alep = 0.094 \pm 0.024 ^{+0.022}_{-0.017}$.

To check the consistency of the measured \Alep{}, the sample is divided in positive and negative charged
leptons, and in the electrons and muons channels.
\Alep{} is measured separately in all the four sub-samples using the same procedure as for the combined measurement.
The results, shown in table \ref{tab:splitresults}, are all consistent with the measured value in the combined sample.
\begin{table}
\centering
\caption{Measurement of \Alep{} in $l^+$, $l^-$, electrons and muons sub-samples.}
\label{tab:splitresults}       
\begin{tabular}{ll}
\hline
Sub-sample & measured \Alep{} \\
\hline
\hline
Positive  & $0.125 ^{+0.041}_{-0.041}$ \\
Negative  & $0.063 ^{+0.046}_{-0.042}$ \\
\hline
Electrons & $0.062 ^{+0.052}_{-0.049}$ \\
Muons     & $0.119 ^{+0.039}_{-0.037}$ \\
\hline
\end{tabular}
\end{table}

The measured value of \Alep{} is in good agreement with the D0 measurement $\Alep = 0.118 \pm 0.032$, and
can be compared to the fixed order NLO QCD+EW prediction $0.038 \pm 0.003$ \cite{Bernreuther:2012sx}.


%
\bibliography{thebib}

\begin{thebibliography}{14}

\bibitem{Aaltonen:2011kc}
T.~Aaltonen et~al. (CDF Collaboration), Phys.Rev. \textbf{D83}, 112003 (2011),
  \texttt{1101.0034}

\bibitem{Abazov:2011rq}
V.M. Abazov et~al. (D0 Collaboration), Phys.Rev. \textbf{D84}, 112005 (2011),
  \texttt{1107.4995}

\bibitem{Bernreuther:2012sx}
W.~Bernreuther, Z.G. Si, Phys.Rev. \textbf{D86}, 034026 (2012),
  \texttt{1205.6580}

\bibitem{Ahrens:2011uf}
V.~Ahrens, A.~Ferroglia, M.~Neubert, B.D. Pecjak, L.L. Yang, Phys.Rev.
  \textbf{D84}, 074004 (2011), \texttt{1106.6051}

\bibitem{Hollik:2011ps}
W.~Hollik, D.~Pagani, Phys.Rev. \textbf{D84}, 093003 (2011), \texttt{1107.2606}

\bibitem{Campbell:2012uf}
J.M. Campbell, R.K. Ellis (2012), \texttt{1204.1513}

\bibitem{Brodsky:2012ik}
S.J. Brodsky, X.G. Wu, Phys.Rev. \textbf{D85}, 114040 (2012),
  \texttt{1205.1232}

\bibitem{Abazov:2012oxa}
V.M. Abazov et~al. (D0 Collaboration) (2012), \texttt{1207.0364}

\bibitem{Falkowski:2012cu}
A.~Falkowski, M.L. Mangano, A.~Martin, G.~Perez, J.~Winter (2012),
  \texttt{1212.4003}

\bibitem{Berger:2012tj}
E.L. Berger, Q.H. Cao, C.R. Chen, H.~Zhang (2012), \texttt{1209.4899}

\bibitem{Sjostrand:2006za}
T.~Sjostrand, S.~Mrenna, P.Z. Skands, JHEP \textbf{0605}, 026 (2006),
  \texttt{hep-ph/0603175}

\bibitem{Mangano:2002ea}
M.L. Mangano, M.~Moretti, F.~Piccinini, R.~Pittau, A.D. Polosa, JHEP
  \textbf{0307}, 001 (2003), \texttt{hep-ph/0206293}

\bibitem{Frixione:2007nw}
S.~Frixione, P.~Nason, G.~Ridolfi, JHEP \textbf{0709}, 126 (2007),
  \texttt{0707.3088}

\bibitem{Alwall:2007st}
J.~Alwall, P.~Demin, S.~de~Visscher, R.~Frederix, M.~Herquet et~al., JHEP
  \textbf{0709}, 028 (2007), \texttt{0706.2334}

\end{thebibliography}
%
%
%
%

\end{document}